\documentclass[aip,graphicx,amsmath,amssymb,reprint]{revtex4-2}

\usepackage{graphicx}
\usepackage{dcolumn}
\usepackage{bm}

\usepackage[utf8]{inputenc}
\usepackage[T1]{fontenc}
\usepackage{mathptmx}
\usepackage{siunitx}
\usepackage{xspace}
\usepackage{braket}
\usepackage{sidecap}
\usepackage{float}

\usepackage{todonotes}

\usepackage{hyperref}

\DeclareSIUnit{\samples}{S}
\DeclareSIUnit{\bit}{Bit}
\DeclareSIUnit{\Vpp}{Vpp}

\newcommand*{\eq}{Eq.~}
\newcommand*{\fig}{FIG.~}
\newcommand*{\sect}{Sect.~}

\newcommand{\ie}{i.e.\ }
\newcommand{\eg}{e.g.\ }
\newcommand{\cf}{cf.\ }

\newcommand{\hub}{Institut für Physik, Humboldt-Universität zu Berlin, 12489 Berlin, Germany}
\newcommand{\quartiq}{QUARTIQ GmbH, Rudower Chaussee 29, 12489 Berlin, Germany}
\newcommand{\fbh}{Ferdinand-Braun-Institut gGmbH, Leibniz-Institut für Höchstfrequenztechnik, 12489 Berlin, Germany}

\begin{document}

\preprint{AIP/123-QED}

\title[Linien: A versatile, user-friendly, open-source FPGA-based tool for frequency stabilization and spectroscopy parameter optimization]{Linien: A versatile, user-friendly, open-source FPGA-based tool for frequency stabilization and spectroscopy parameter optimization}

\author{B. Wiegand}
\email{benjamin.wiegand@physik.hu-berlin.de}
\affiliation{\hub}

\author{B. Leykauf}
\affiliation{\hub}

\author{R. Jördens}
\affiliation{\quartiq}

\author{M. Krutzik}
\affiliation{\hub}
\affiliation{\fbh}

\date{\today}

\begin{abstract}
  \begin{center}
  The following article has been accepted by \emph{Review of Scientific Instruments}.\\
  After it is published, it will be found at \url{https://doi.org/10.1063/5.0090384}
  \end{center}
We present a user-friendly and versatile tool for laser frequency stabilization. Its main focus is spectroscopy locking, but the software is suitable for lock-in techniques in general as well as bare PID operation. Besides allowing for sinusoidal modulation (up to \SI{50}{\MHz}), triangular ramp scanning, IQ demodulation (1 $f$ to 5 $f$), IIR and PID filtering, \emph{Linien} features two different algorithms for automatic lock point selection; one of them performs time-critical tasks completely on FPGA. \emph{Linien} is capable of autonomously optimizing spectroscopy parameters by means of machine learning and can measure the error signal's power spectral density. The software is built in a modular way, providing both a graphical user interface as well as a Python scripting interface. It is based on the RedPitaya STEMLab platform but may be ported to different systems.
\end{abstract}

\maketitle


\section{Introduction}

Atomic, molecular, and optical (AMO) physics experiments rely on laser frequencies that are stabilized by means of spectroscopic methods. This requires an active feedback loop that is typically provided by a proportional-integral-derivative (PID) controller. While this functionality was traditionally implemented using analog circuits, nowadays digital devices are often used as they are easy to reconfigure or reprogram while providing sufficient control bandwidth \cite{marceluda, neuhaus2017pyrpl, jorgensen2016simple, spindeldreier2018fpga, roy2019fpga, stimpson2019open, pomponio2020fpga, preuschoff2020digital, yu2018performance, tourigny2018open, schwettmann2011field, yang2012low, leibrandt2015open, xu2012digital, sparkes2011scalable,kobayashi2019relocking, hannig2018highly}. While microcontroller-based devices exhibit a limited feedback bandwidth in the $\SI{100}{\kilo\hertz}$ range, field-programmable gate arrays (FPGA) allow for reaching the $\SI{}{\mega\hertz}$ regime. Furthermore, digital feedback controllers may be implemented using compact and affordable off-the-shelf components: the RedPitaya STEMlab platform for example combines a Xilinx Zynq 7010 integrated circuit core featuring a dual-core ARM Cortex-A9 CPU and an FPGA with 10-bit or 14-bit analog-to-digital converters (ADCs) and digital-to-analog converters (DACs) in a small form factor.

For this platform, we developed \emph{Linien}: an open-source tool for laser frequency stabilization with a strong focus on user friendliness. On the one hand, the software is designed to work out of the box for most spectroscopy setups. On the other hand, it is highly configurable in order to suit different experimental needs. Furthermore, it features algorithms like automatic lock point selection as well as autonomous parameter optimization that are intended to make lab life easier. To the best of our knowledge, comparable algorithms have not yet been presented in similar software for the RedPitaya STEMlab platform\cite{preuschoff2020digital,marceluda,stimpson2019open,tourigny2018open, neuhaus2017pyrpl}.

In this paper, we describe the system design of \emph{Linien} and present exemplary use cases for this software.

\section{System design}

\begin{figure*}
	\includegraphics[width=.9\linewidth,trim={0 0cm 0 0cm},clip]{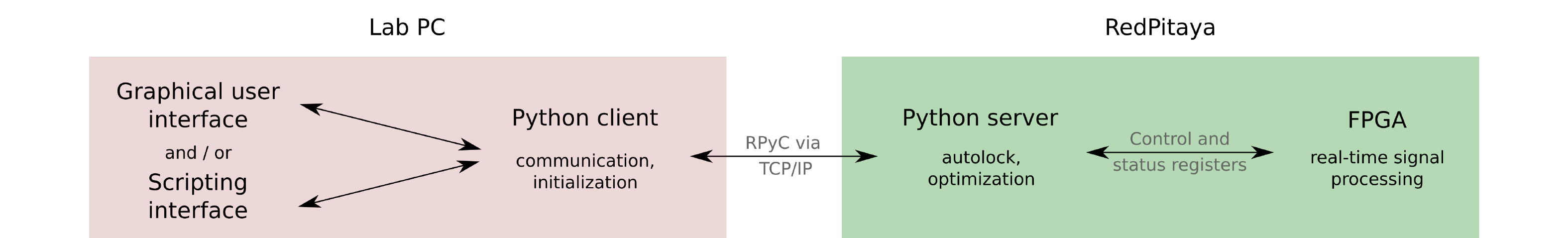}
	\caption[]{Software architecture of \emph{Linien}. The server component (green) is capable of operating autonomously and may be controlled by one or more clients (red).}
	\label{fig:architecture}
\end{figure*}

\emph{Linien} runs on the RedPitaya STEMLab platform (125-10 or 125-14 version) and follows a server-client architecture, with the server component being capable of operating autonomously on the RedPitaya. It is itself composed of two submodules: the FPGA firmware for real-time signal processing and a Python module that implements higher-level functions like the autolock (\cf \sect \ref{sec:autolock}) and provides a communication interface.

If desired, the server may be controlled by one (or even multiple) clients. The core component of the client is a Python library that communicates with the server via TCP/IP using the Python package RPyC\cite{rpyc}, which allows for seamless remote procedure calls. \emph{Linien}'s client library may be used directly in Python programs or may be controlled by means of a graphical user interface (GUI). The overall architecture of \emph{Linien} is depicted in \fig \ref{fig:architecture}.

\begin{figure}[h]
	\includegraphics[width=\linewidth,trim={0 0cm 0 0cm},clip]{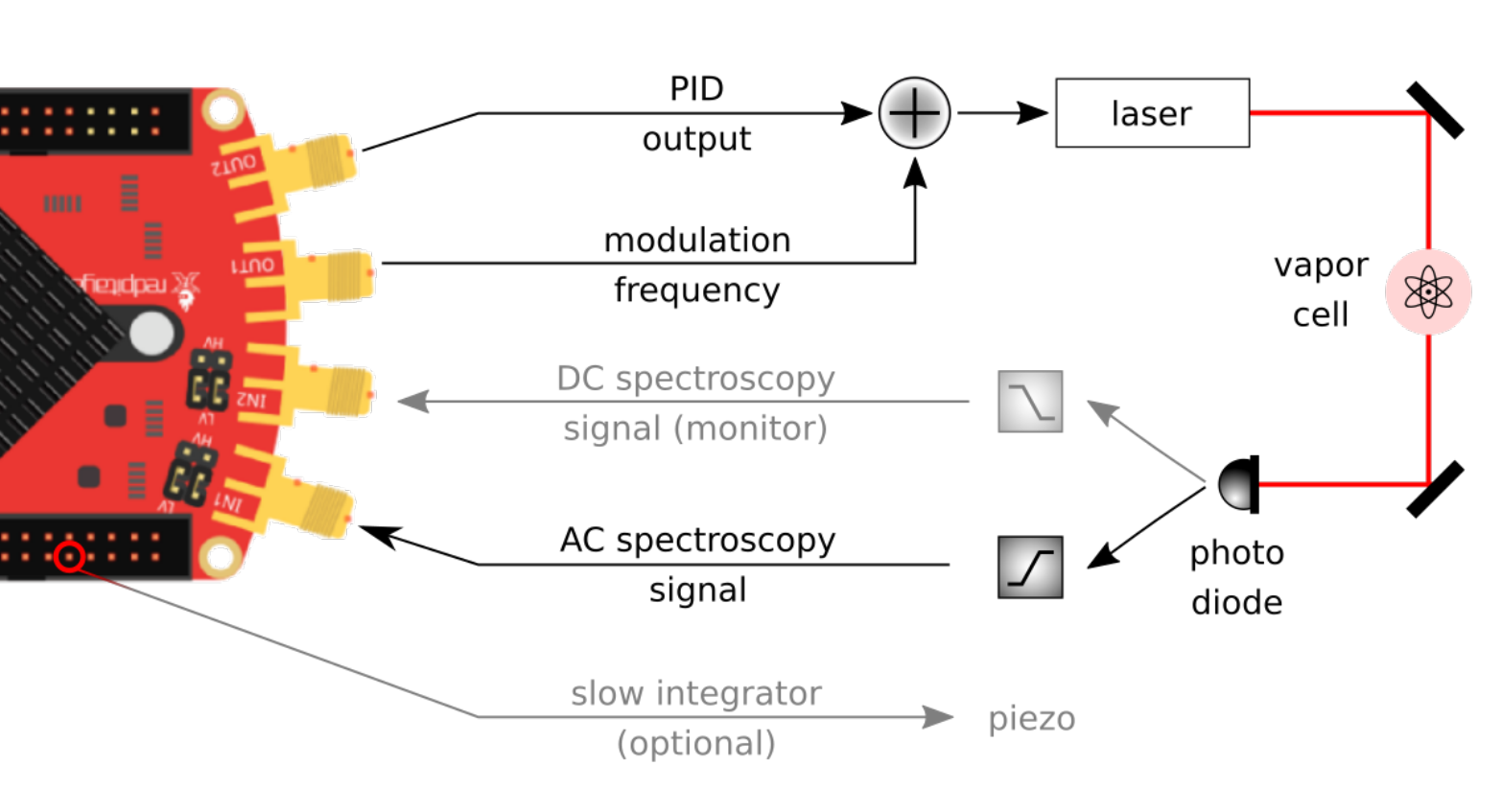}
	\caption[]{A sample setup for stabilizing a laser frequency to an atomic transition by means of spectroscopy. Laser light passes an atomic ensemble and is picked up by a photo diode. The signal's AC part is fed to a fast input of RedPitaya, demodulated by Linien and used as the error signal for the PID. After filtering, the output signal closes the feedback loop by controlling the laser current. Note that instead of directly applying the modulation frequency to the laser current, a dedicated modulator like an acousto-optic or an electro-optic modulator may be used.}
	\label{fig:setup}
\end{figure}

As a core objective of \emph{Linien} is ease of use, a standalone executable for Linux and Windows is provided\footnote{https://github.com/linien-org/linien} that runs on the user's computer and is capable of installing all the requirements on a RedPitaya device without manual intervention. A sample setup for spectroscopy locking using \emph{Linien} is shown in \fig \ref{fig:setup}.

\subsection{FPGA architecture}

The code of the FPGA firmware is mainly written in Python and transpiled to Verilog using migen\cite{migen}, a toolbox for designing and simulating digital circuits. \emph{Linien}'s FPGA firmware is based on the heritage of redpid\cite{redpid}.

\begin{figure*}
	\includegraphics[width=0.80\linewidth,trim={0 0cm 0 0cm},clip]{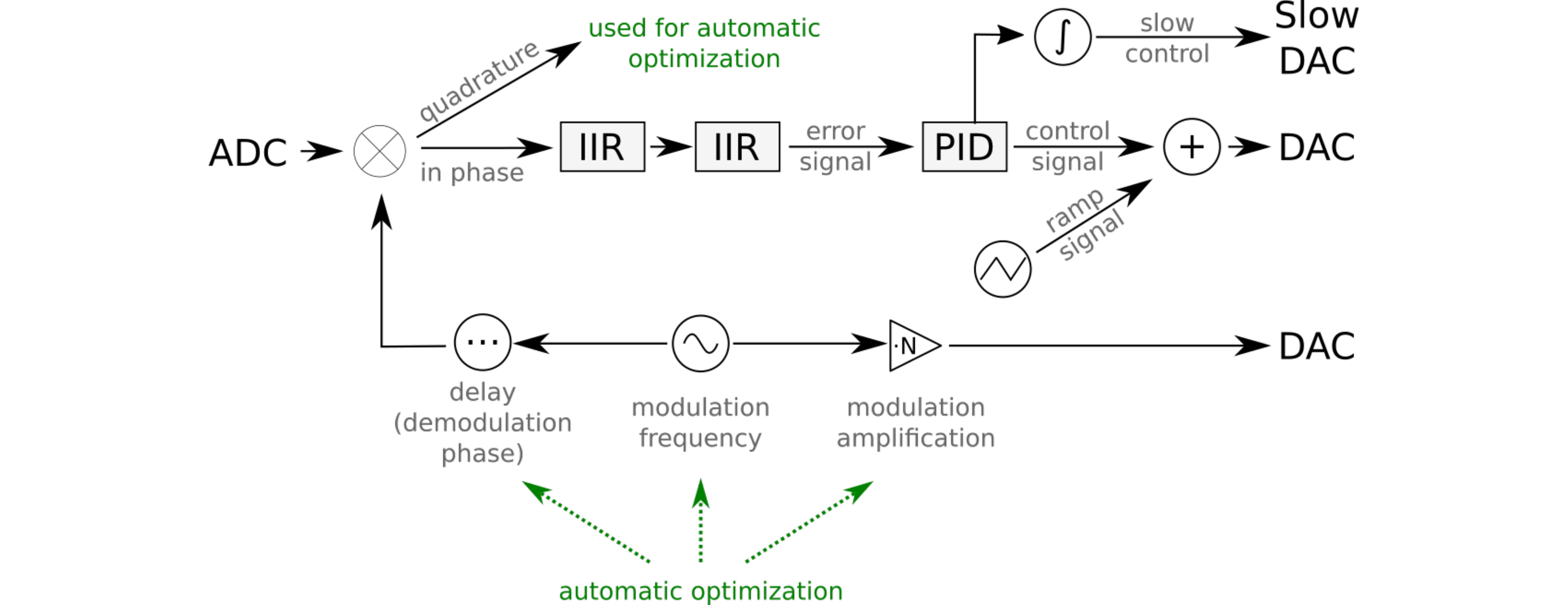}
	\caption[]{Simplified schematic of the signal flow inside the FPGA. Note that this figure depicts the default setup; different modes of operation are available. Automatic optimization, depicted in green, employs machine learning on the CPU in order to find optimized spectroscopy parameters with a higher signal-to-noise ratio.}
	\label{fig:fpga}
\end{figure*}

\fig \ref{fig:fpga} gives a simplified overview of the signal flow for a feedback loop of a spectroscopy lock: the spectroscopy signal is picked up using the RedPitaya's ADCs at a rate of $\SI{125}{\mega\hertz}$, mapping the input range of $\pm\SI{1}{\volt}$ to a $\SI{14}{\bit}$ integer. Starting at this point, all signal processing takes place using integers with a width of \SI{25}{\bit} in order to minimize the impact of numerical errors.

In order to demodulate the spectroscopy signal, the CORDIC algorithm\cite{cordic} is employed. For this purpose, the modulation frequency (1 $f$) or a multiple of it (up to 5 $f$) may be used. The in-phase component is processed by two consequent infinite impulse response (IIR) filters with user-defined parameters, yielding the \emph{error signal}.

The next building block is a proportional-integral-derivative (PID) with user-defined coefficients.
The output of the PID filter is the \emph{control signal}. The RedPitaya's digital-to-analog converter (DAC) with a sampling rate of $\SI{125}{\mega\samples\per\second}$ maps this signal to $\pm\SI{1}{\volt}$ and enables the user to close the feedback loop. In order to increase the capture and control range, an additional integrator with configurable strength may be used that is available at a slow analog output pin ($\SI{0}{\volt} - \SI{1.8}{\volt}$). This signal is provided by first order delta-sigma modulation \cite{deltasigma} of this port which means that additional analog low-pass-filtering is recommended.

It is worth noting that the signal flow presented above is not the only mode of operation: the output ports of control signal, ramp signal and modulation signal are variable.
If the error signal is provided externally (for example by a phase-frequency detector), bare PID operation with no demodulation at all can be used for offset locking (\cf \sect \ref{sec:bare-pid}).
As the ramp's output channel is variable, a piezo controller may be used to scan the laser frequency.
Furthermore, dual-channel operation (\ie demodulation of both fast input channels) for combined frequency modulation spectroscopy (FMS, \cite{fms}) and modulation transfer spectroscopy (MTS, \cite{mts}) is supported\cite{combined-fms-mts}.

\subsection{Autolock for spectroscopy stabilization}\label{sec:autolock}
In order to stabilize laser frequency to an atomic transition, Doppler-free saturation spectroscopy methods are typically used for high precision experiments. When turning on the laser stabilization, care has to be taken to actually pick the correct zero crossing as various spectral features may be visible. In order to help with that, \emph{Linien} features two \emph{autolock algorithms} that let the user select a target transition in the graphical user interface and then automatically lock to this zero crossing: \emph{simple autolock} (\sect \ref{sec:fast-autolock}) and \emph{jitter tolerant autolock} (\sect \ref{sec:jitter-tolerant-autolock}). By default, \emph{Linien} chooses a suitable algorithm depending on the magnitude of laser frequency jitter it detects. Please note that the purpose of the algorithms presented below is not automatic relocking after loss of lock similar to implementations in several other projects \cite{roy2019fpga,kobayashi2019relocking,marceluda}. Instead, the algorithms serve the purpose of facilitating initial lock point selection.

\subsubsection{Simple autolock}\label{sec:fast-autolock}

The \emph{simple autolock} is similar to what has already been presented in literature\cite{marceluda}: the user selects the desired line from the spectrum that is continuously recorded while the laser frequency is sweeped. Then, another spectrum is recorded which is used to determine the current laser frequency relative to the reference spectrum. This is achieved by maximizing the correlation between the two spectra. Using this information, the position of the ramp at which the target transition is expected is calculated; the FPGA's locking module is programmed to start PID operation once this ramp position is reached.

While this algorithm works well for many setups, it has a central flaw: if the laser exhibits strong frequency jitter, the targeted line may shift significantly during the runtime of the algorithm, potentially causing the lock to fail or to target a different zero crossing. This is due to the time it takes to record a spectrum, to transfer it from FPGA (where it is recorded) to CPU, to calculate the correlation function on the CPU and to wait until the ramp is at the correct position for engaging the lock.

\subsubsection{Jitter tolerant autolock}\label{sec:jitter-tolerant-autolock}

\emph{Linien} features a novel kind of autolock algorithm that performs time-critical tasks directly on the FPGA instead of relying on the CPU. This eliminates the delay due to the time it takes to transfer a spectrum from FPGA to CPU altogether. Furthermore, the algorithm is designed to analyze the spectrum in real-time (\ie while the ramp is running) and to turn on the lock at the correct position. This means that slow ramp speeds don't delay the start of the lock.

It is worth noting that this method is not based on calculation of the correlation function, as doing so on the FPGA in real-time would be a challenging task. Instead, a different approach was chosen, tailored to the functionalities that are easily implemented on an FPGA.

The starting point for the algorithm is again a reference spectrum with respect to which the user selects the target transition. Then, several additional spectra are recorded. The part of the algorithm that runs on the CPU now analyzes all these spectra, trying to understand what features are characteristic (this process is described in detail in \sect \ref{sec:jitter-tolerant-autolock:detail}). As a result, a description of the predominant features of the spectrum as well as instructions on how to find the lock point with respect to these features is obtained. This information is sent to the FPGA where it is used to autonomously analyze the spectrum in real-time (\ie while the ramp is running) and to turn on the lock at the right point.

The main advantage of this method is that time-critical operations are executed directly on the FPGA. This allows for starting the lock at exactly the right position, with no delays being involved. As a result, even a laser that exhibits high frequency jitter can be locked reliably.

\paragraph{Detailed description of feature recognition}\label{sec:jitter-tolerant-autolock:detail}

As explained above, multiple spectra are recorded. The algorithm running on the CPU analyzes them, tries to detect the prominent features, and generates a description that is used by the FPGA to turn on the lock at the right position of the ramp.

\begin{figure}[h]
	\includegraphics[width=\linewidth,trim={0 0cm 0 0cm},clip]{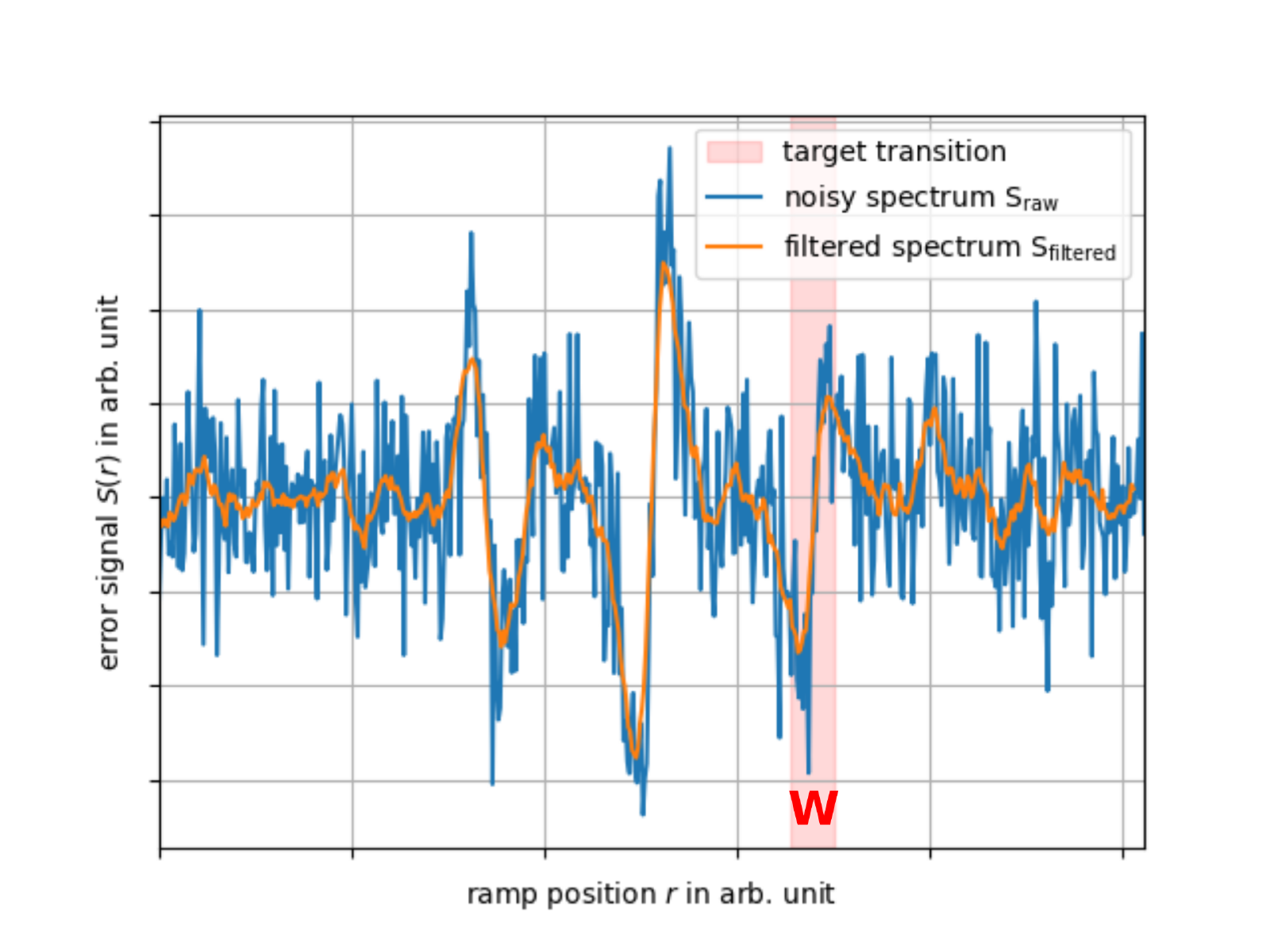}
	\caption[]{A simulated noisy test spectrum before (blue) and after (orange) filtering according to \eq \ref{eq:filtering}. For easy comparison of the spectra, the filtered spectrum was shifted to the left by $0.5\,W$. The red region with width $W$ is the target transition (normally selected by the user).}
	\label{fig:spectrum-filtering}
\end{figure}

As a first step in implementing this algorithm, we had to find a method for noise reduction that can be calculated on a CPU (for feature description) and on an FPGA (for real-time feature recognition). For this, we make use of the information that the user provides by selecting a target transition: we know the linewidth $W$ of the feature of interest by analyzing the peaks in the user-selected region. Hence, we low-pass filter the spectrum $S_\text{raw}(r)$ according to
\begin{align}
	\label{eq:filtering}
	S_\text{filtered}(r) = \frac{1}{W} \cdot \int_{r-W}^r S_\text{raw}(r') \,dr'\ \ ,
\end{align}
with $r$ being the ramp position. This is exemplarily shown in \fig \ref{fig:spectrum-filtering} with a simulated spectrum.

\begin{figure}[h]
	\includegraphics[width=\linewidth,trim={0 0cm 0 0cm},clip]{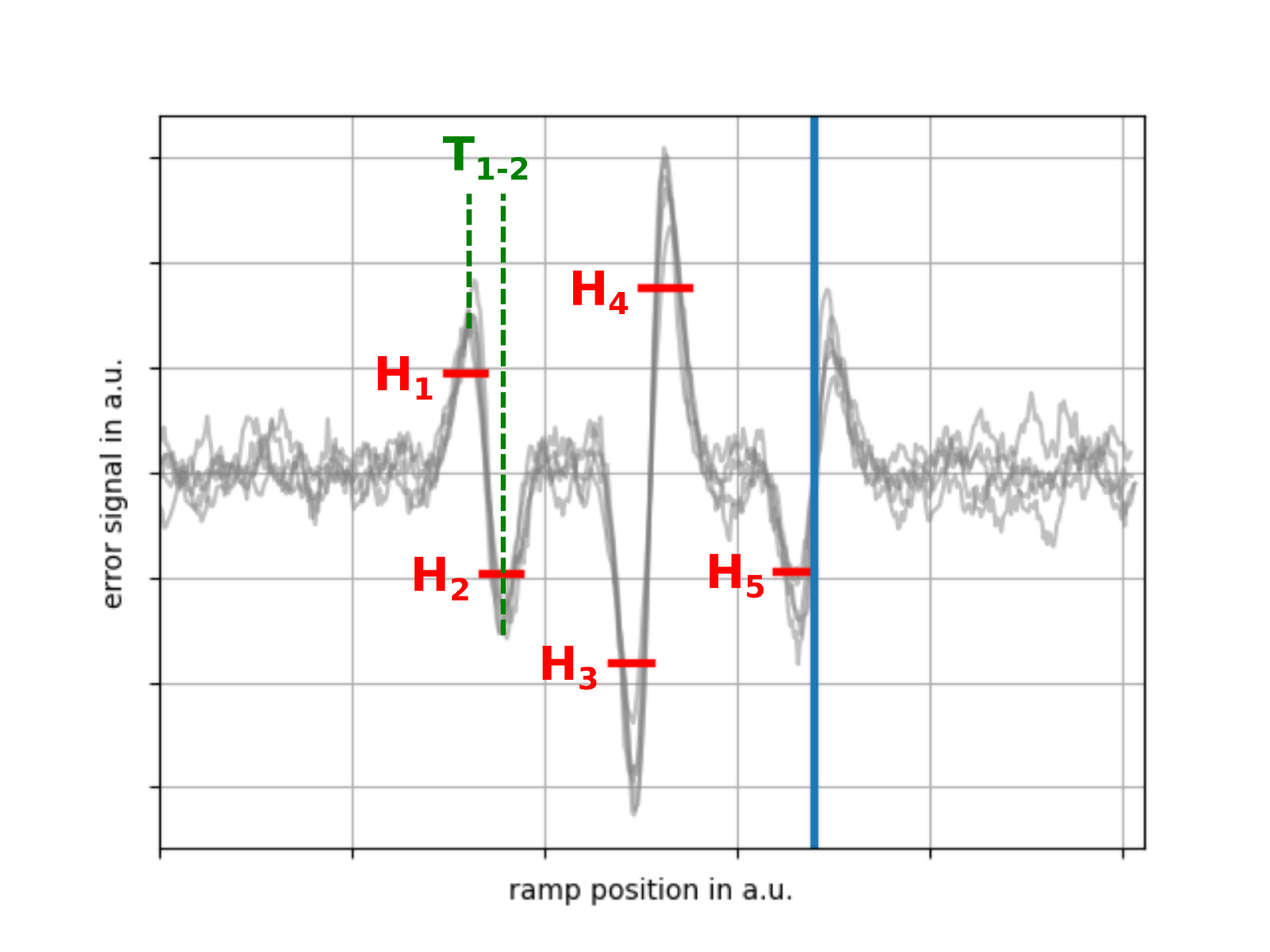}
	\caption[]{Demonstration of the jitter-tolerant autolock's principle of operation: given 5 synthetic noisy test spectra that were filtered according to \eq \ref{eq:filtering} (grey), the algorithm detects peaks that are present in all the spectra. For each peak, a threshold is determined (red) that allows to recognize this feature in combination with the delay between the features ($T_\text{1-2}$, green in this example).The blue line shows the lock position as determined by the autolock. For details on the algorithm see the main text.}
	\label{fig:autolock}
\end{figure}

The second step is to detect peaks in the filtered spectrum (\fig \ref{fig:autolock}). The crucial part here is that we only take into account peaks that are found in all the recorded spectra, which allows us to neglect random fluctuations. In a realistic scenario, the determined peak heights differ between the spectra though. Hence, we lower the threshold for peak detection until we get a description that matches all the recorded spectra (including a small safety margin). Altogether, we have generated a set of instructions (\eg \emph{after a peak of height $H_1$ and another peak of height $H_2$ wait for a time $T_\text{1-2}$ engaging the lock}), explaining how to find the lock point.

In the final step, we use this information on the FPGA in order to find the target transition in real-time (\ie while the ramp is running) and to close the feedback loop at the correct position.

\paragraph{Verification by simulation}

As the FPGA code is written in Python with migen, it is easy to simulate it. In order to verify the functionality of the jitter-tolerant autolock, a test suite that runs it on various different sample spectra with simulated noise and laser jitter was used. In $100$ runs, the algorithm always succeeded in finding the correct lock position.


\subsection{Optimization of spectroscopy signal using machine learning}

For atomic spectroscopy techniques, the strength of the spectroscopy signal and its signal-to-noise ratio depend strongly on the parameters used (\ie modulation frequency, modulation amplitude and demodulation phase). When stabilizing a laser to an atomic transition it is crucial that small deviations in frequency lead to a strong error signal, allowing the feedback loop to counteract effectively. In other words: in order to obtain a good frequency stability, the steepness of the transition's slope at the zero crossing should be as high as possible.

Therefore, \emph{Linien} features automatic optimization of the spectroscopy signal using machine learning. We chose the Python implementation\cite{pycma} of the Covariance matrix adaptation evolution strategy (CMA-ES, \cite{cmaes}) as underlying algorithm, because this technique has been shown to exhibit fast convergence for a variety of blackbox functions\cite{cmaes-performance} and has moderate computational cost.

By employing IQ demodulation, we effectively reduce the size of the parameter space to 2 ($U_\text{mod}$ and $f_\text{mod}$): the machine learning algorithm records spectra for different sets of modulation frequency and amplitude. As the signal's in-phase and quadrature components are orthogonal, the optimized demodulation phase for each set of parameters can be calculated. This method speeds up the optimization process compared to a direct optimization of all three modulation / demodulation parameters.

\begin{figure}[h]
	\includegraphics[width=\linewidth,trim={0 0cm 0 0cm},clip]{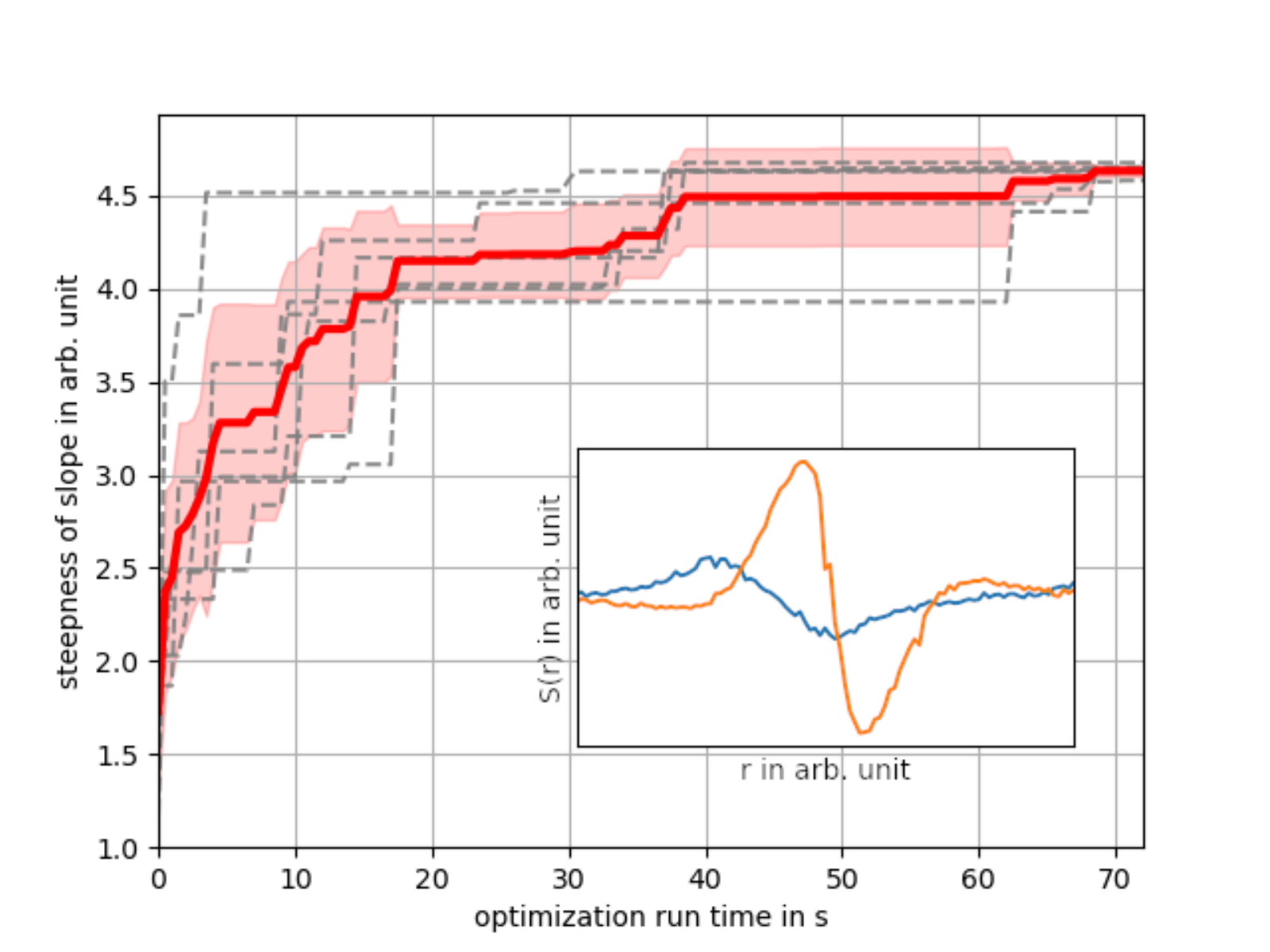}
	\caption[]{Exemplary runs of automatic optimization of spectroscopy parameters (dashed grey). The red line depicts the mean steepness of the transition's slope, the light red area shows the standard deviation. Initial parameters ($f_\text{mod}=\SI{8}{\mega\hertz}$, $U_\text{mod, pp}=\SI{1}{\volt}$) and boundaries ($ \SI{6}{\mega\hertz} \leq f_\text{mod} \leq \SI{10}{\mega\hertz}$, $\SI{1}{\volt} \leq U_\text{mod, pp} \leq \SI{2}{\volt}$) were the same for all runs. Optimized parameters ($f_\text{mod} \approx \SI{8.6}{\mega\hertz}$, $U_\text{mod, pp} \approx \SI{1.9}{\volt}$) were found after roughly a minute. The inlay shows (in arb. unit) the MTS spectrum before (blue) and after (orange) optimization. In this plot, $r$ is the ramp position and $S(r)$ the corresponding spectroscopy signal.}
	\label{fig:optimization}
\end{figure}

\fig \ref{fig:optimization} shows several exemplary optimization runs conducted with the rubidium MTS spectroscopy module of the mobile atom interferometer GAIN\cite{gain}. In these experiments, the \textsuperscript{85}Rb $F=3 \rightarrow F'=4$ transition's steepness was optimized.

In our tests, the algorithm was capable of optimizing the steepness of the spectroscopy signal at the zero crossing within approximately one minute, corresponding to roughly $100$ iterations. On our platform, with a real spectrum, this corresponds to roughly $\SI{1}{\minute}$ (limited by the acquisition time).

\subsection{PSD measurement for PID parameter optimization}

\emph{Linien} allows for measuring the power spectral density (PSD) of the in-loop error signal for analyzing the frequency stability of the locked laser. In order to obtain the PSD, a low-pass filter is applied to the demodulated error signal which reduces alias effects due to limited sampling rate. As the FPGA firmware is only capable of recording and storing 16384 samples at once, multiple chunks are recorded with increasing decimations of the sampling rate. For each of these samples, the power spectral density is calculated either using the LPSD algorithm \cite{lpsd} or Welch's method\cite{welch}. In the end, the chunks are combined to a single PSD curve that may extend from the \si{\mega\hertz} to a user-defined timescale.

\begin{figure}[h]
	\includegraphics[width=\linewidth,trim={0 0cm 0 0cm},clip]{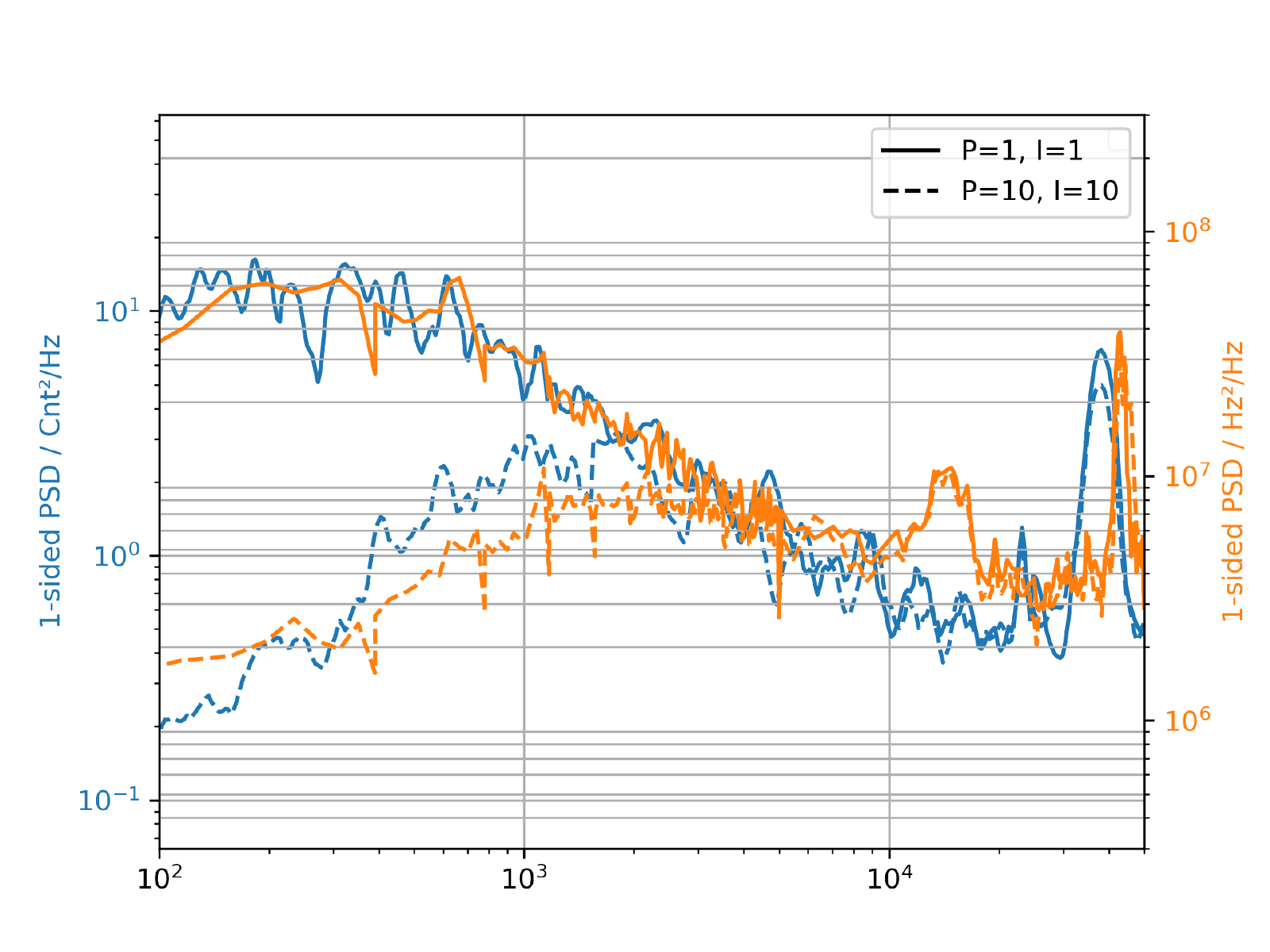}
	\caption[]{Optimization of frequency noise by measurement of the PSD. Blue lines show the PSD of the in-loop error signal of a laser stabilized to an atomic transition using MTS. Solid and dashed lines indicate two different sets of PID parameters used for the locking. The dashed line (higher PID parameters) shows a decreased noise density at lower frequencies. For comparison, the orange lines were acquired by sampling the beat-note frequency between this laser and another laser stabilized to an atomic transition via FMS using a Pendulum CNT-91 counter. Note that this data is not intended to show the lock performance that may be achieved using Linien as it strongly depends on the experimental setup and laser system that are used.}
	\label{fig:psd}
\end{figure}

The measurement of the error signal's PSD is useful for finding suitable PID parameters, as depicted in \fig \ref{fig:psd}. In this graph, the solid lines were recorded with very low PID parameters. In this case, the beat note exhibits strong noise at low frequencies. When increasing the PID parameters (dashed lines) this noise is suppressed.
This improvement in frequency noise can also be observed when measuring the PSD of the beat-note frequency between the laser under test and another frequency stabilized laser (orange lines). As a result, we conclude that the PSD recorded by \emph{Linien} is suited for an optimization of PID parameters.

\subsection{Bare PID operation}\label{sec:bare-pid}

Linien may also be used for bare PID operation. For this purpose, a \emph{fast mode} is available that bypasses demodulation and IIR filters (\cf \fig \ref{fig:fpga}): the ADC output is directly PID-filtered and then connected to the DAC.
This reduces propagation delay between input and output from 320 ns to 125 ns. The remaining lag is mainly due to conversion delays of ADC and DAC. This mode of operation is useful for scenarios with an externally generated error signal and allows for approaching control bandwidths in the low $\SI{}{\mega\hertz}$ regime \cite{yu2018performance}. Using \emph{Linien}'s fast mode, we achieved a control bandwidth of nearly \SI{1}{\mega\hertz} (servo bumps at roughly \SI{900}{\kilo\hertz}) when offset-locking two Toptica DL pro ECDLs with OnSemiconductor MC100EP140 on a custom board being used as phase-frequency detector (PFD).

\section{Conclusion}

We have presented a tool for laser frequency stabilization based on the RedPitaya STEMLab platform that constitutes a low-cost yet versatile solution. It is suited for a wide range of experimental setups and features algorithms for automatic lock point selection, optimization of the spectroscopy signal and PSD measurement for PID parameter optimization. On the one hand, this software is designed as a user-friendly lab tool that facilitates many tedious tasks. On the other hand \emph{Linien}'s algorithms allow to build instruments that operate autonomously without manual intervention \cite{fruby}.

We also presented a novel kind of autolock algorithm that performs time-critical tasks completely on FPGA. This allows to eliminate delays that are inevitable in algorithms that have been previously demonstrated, making this method suitable for signals with strong jitter. Furthermore, it may be used in the future to build systems with minimized size, weight and power (SWaP) budget as no CPU is required for the autolock if a predefined spectrum is to be locked - an FPGA-only board would be sufficient.

\section{Acknowledgements}
This work is supported by the German Space Agency (DLR) with funds provided by the Federal Ministry of Economics and Technology (BMWi) under grant number DLR50WM2066.

We thank Matthias Schoch for his help in debugging the electronics as well as Klaus Döringshoff, Oliver Anton, Julien Kluge and Aaron Strangfeld for their valuable input.
Furthermore, we thank Laurenz Reichl for designing \emph{Linien}'s icon and all contributors on the github project.

\nocite{*}
\bibliography{Paper}

\end{document}